\documentclass[twocolumn,aps,prb,superscriptaddress]{revtex4-1}
\usepackage{amsmath,amssymb,amstext,graphicx,bm}

\begin{document}

\title{Observation of spin-selective tunneling in SiGe nanocrystals}

\author{G. Katsaros}
\altaffiliation{Present address: Institute for Integrative Nanosciences, IFW Dresden,
 Helmholtzstr.\ 20, D-01069 Dresden, Germany, email: g.katsaros@ifw-dresden.de}
\affiliation{SPSMS, CEA-INAC/UJF-Grenoble 1, 17 Rue des Martyrs, 38054 Grenoble Cedex 9, France}

\author{V. N. Golovach}
\affiliation{SPSMS, CEA-INAC/UJF-Grenoble 1, 17 Rue des Martyrs, 38054 Grenoble Cedex 9, France}

\author{P. Spathis}
\affiliation{SPSMS, CEA-INAC/UJF-Grenoble 1, 17 Rue des Martyrs, 38054 Grenoble Cedex 9, France}

\author{N. Ares}
\affiliation{SPSMS, CEA-INAC/UJF-Grenoble 1, 17 Rue des Martyrs, 38054 Grenoble Cedex 9, France}

\author{M. Stoffel}
\affiliation{Institute for Integrative Nanosciences, IFW Dresden,
 Helmholtzstr.\ 20, D-01069 Dresden, Germany}

\author{F. Fournel}
\affiliation{CEA, LETI, MINATEC, F38054 Grenoble, France}

\author{O. G. Schmidt}
\affiliation{Institute for Integrative Nanosciences, IFW Dresden,
 Helmholtzstr.\ 20, D-01069 Dresden, Germany}

\author{L. I. Glazman}
\affiliation{Department of Physics, Yale University, New Haven, Connecticut 06520, USA}

\author{S. De Franceschi}
\affiliation{SPSMS, CEA-INAC/UJF-Grenoble 1, 17 Rue des Martyrs, 38054 Grenoble Cedex 9, France}

\pacs{73.23.Hk; 71.70.Ej; 73.63.Kv}
\date{\today{}}

\begin{abstract}
Spin-selective tunneling of holes in SiGe nanocrystals contacted by normal-metal leads is reported.
The spin selectivity arises from an interplay of the orbital effect of the magnetic field with the strong
spin-orbit interaction present in the valence band of the semiconductor.
We demonstrate both experimentally and theoretically that spin-selective tunneling in semiconductor nanostructures
can be achieved without the use of ferromagnetic contacts.
The reported effect, which relies on mixing the light and heavy holes,
should be observable in a broad class of quantum-dot systems formed in semiconductors with a degenerate valence band.
\end{abstract}

\maketitle

The spin-orbit interaction (SOI) has become of central interest in the past years~\cite{Winkler}, because it enables an all-electrical manipulation of the spin. In
the field of spin qubits, one of us~\cite{Golovach} suggested the electrical
control of localized spins by means of the electric-dipole spin
resonance, and this scheme has been successfully used for spin rotations of electrons
in quantum dots (QDs)~\cite{Nowack,Stevan}. Already much earlier,
Datta and Das~\cite{Datta} proposed a semiconductor transistor that would
operate through a gate-controlled spin precession, mediated by the SOI.
In this type of spin transistor,
spin-polarized electrons are injected into the
semiconductor from a ferromagnetic (FM) contact.
The realization of an efficient spin injection
has proven to be a difficult task~\cite{Zutic,Rashba}.
Only recently, high spin-injection efficiencies were reported for FM contacts to semiconductors~\cite{Koo,Appelbaum,Jonker1}.
In nanostructures, however,
experimental evidence of spin injection is not as strong and clear~\cite{Tsukagoshi,Sahoo,Hamaya1,Zwanenburg,Liu}.
Here we show that the SOI in the valence band, quantified by the spin-orbital splitting $\Delta_{\rm SO}$,
provides an alternative way to obtain spin-selective tunneling without requiring FM electrodes.

At cryogenic temperatures, transport through QDs is dominated by the Coulomb blockade (CB) effect.
In the CB regime, single-hole transport is suppressed and electrical conduction is due to second-order cotunneling (CT) processes ~\cite{Franceschi}.
We consider here the case of a QD with an odd number of holes and a spin-doublet ground state.
A magnetic field, $B$, lifts the spin degeneracy by
the Zeeman energy $E_{Z} = g \mu_B B$, where $g$ and $\mu_B$ are the hole g-factor and Bohr magneton, respectively.
Once the bias voltage across the QD exceeds the Zeeman energy, $\left|eV\right|>E_{Z}$,
the inelastic CT processes can flip the QD spin, leaving the QD in the excited spin state;
hereinafter $e$ is the elementary charge ($e>0$).
The onset of spin-flip inelastic CT manifests itself as a step in the differential conductance, $G = dI/dV$, at $eV = \pm E_{Z}$ ~\cite{Kogan}.

\begin{figure}
\includegraphics[width=0.531\columnwidth]{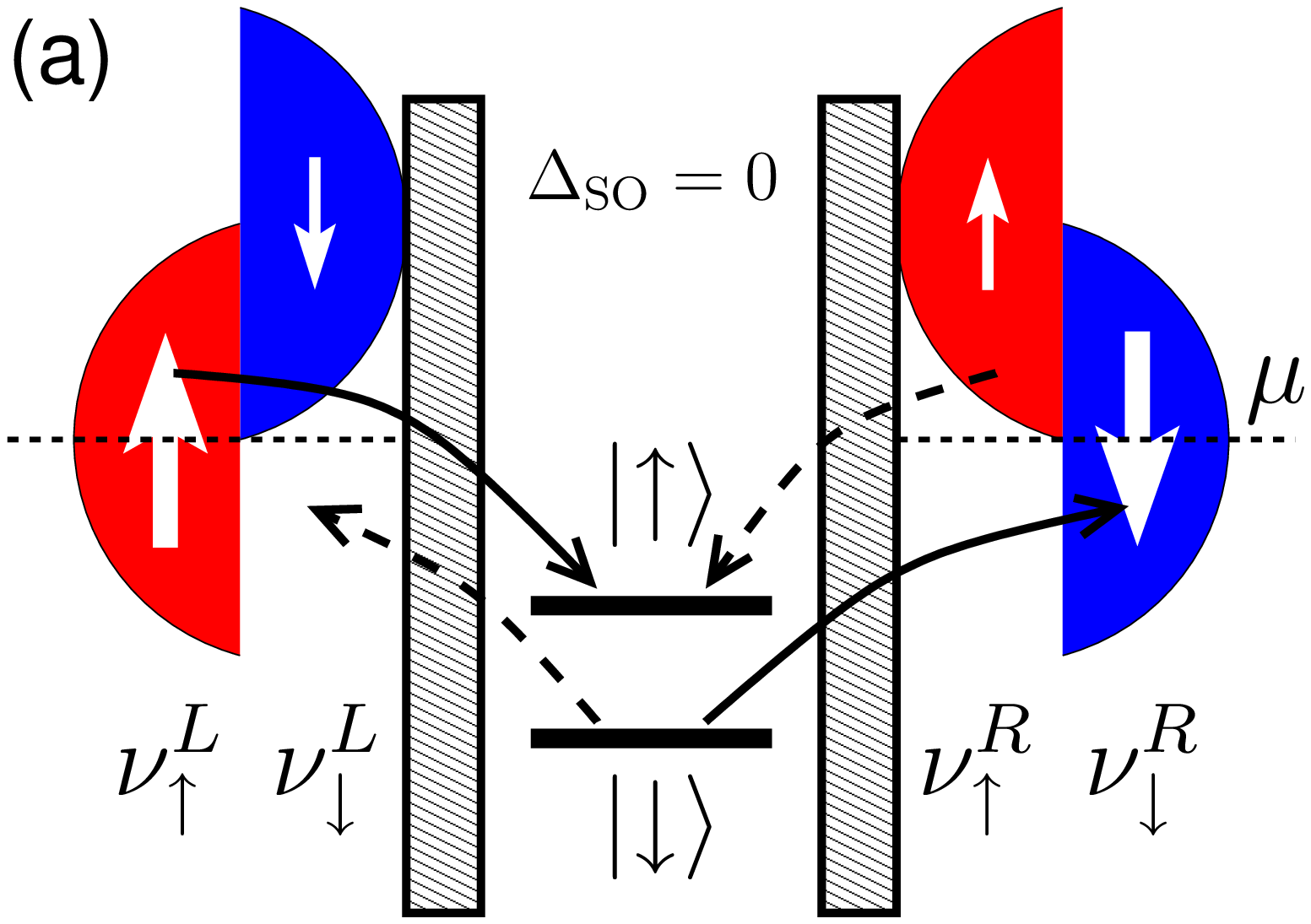}\hfill
\includegraphics[width=0.429\columnwidth]{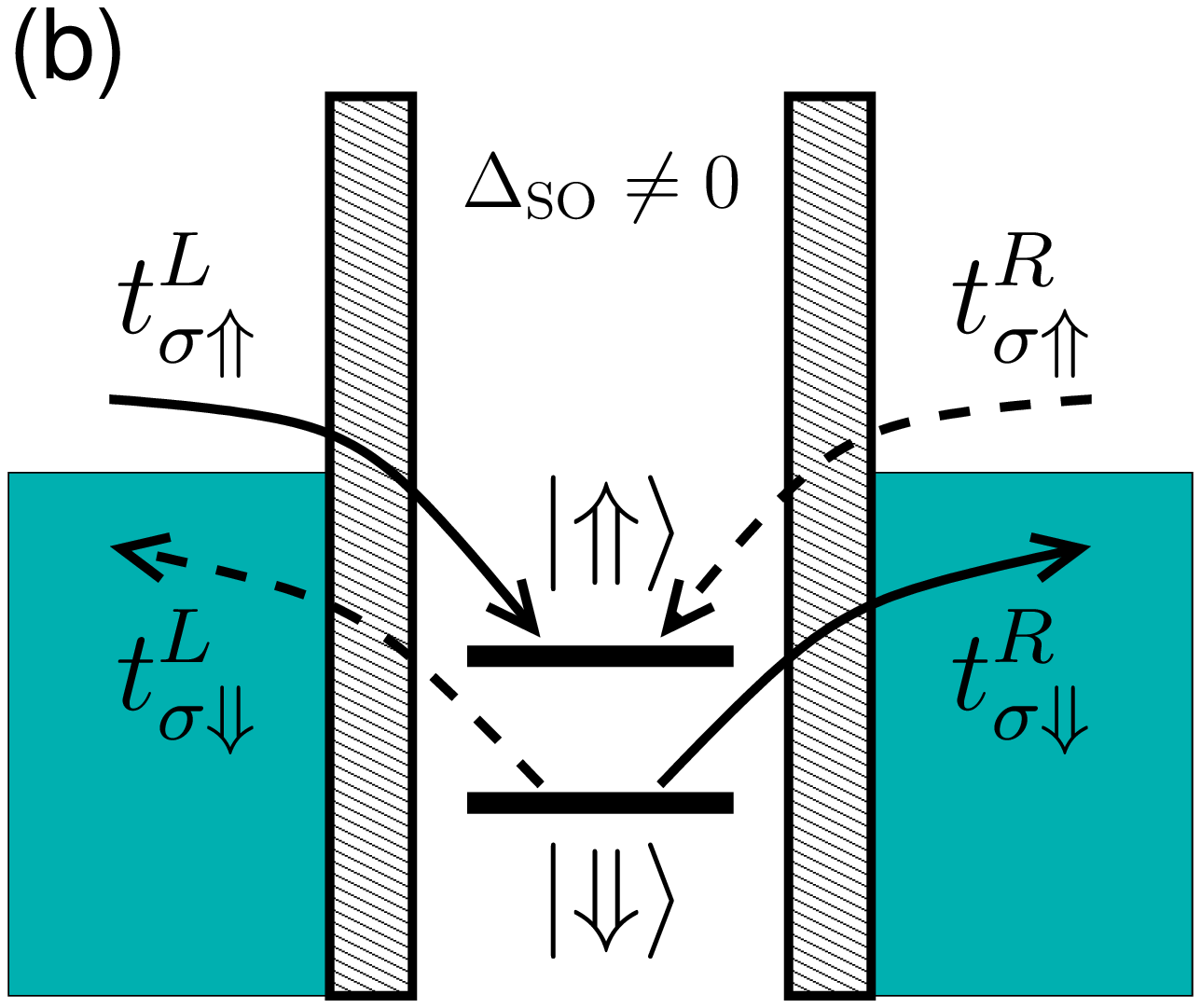}
\caption{
(color online) Spin-selective tunneling in
(a) a QD coupled to FM leads and
(b) a QD with SOI coupled to non-magnetic leads. The solid (dashed) arrows indicate the tunneling processes
involved in the inelastic CT for forward (reverse) biasing, with solid (dashed) arrows representing stronger (weaker) tunnel rates.
In both (a) and (b), the tunnel rate, $\Gamma\equiv\pi\nu\left|t\right|^2$, differs
for each Zeeman sublevel of the QD.
In setup (a), it is the density of states $\nu$ that brings about the spin
selectivity of the tunneling.
In setup (b), the spin selectivity is caused by the tunneling amplitude $t$,
which is sensitive to the spinor wave functions at the point of tunneling.
In the valence band, for energies $\lesssim\Delta_{\rm SO}$,
the $B$-field efficiently makes $\Gamma$ spin-dependent
by affecting the mixing between heavy and light holes.
Since the inelastic CT current is proportional
to $\Gamma_\Uparrow^L\Gamma_\Downarrow^R$ for the forward bias
and to $\Gamma_\Uparrow^R\Gamma_\Downarrow^L$ for the reverse bias,
an asymmetric $G(V)$ is expected whenever
$\Gamma_\Uparrow^L\Gamma_\Downarrow^R \neq \Gamma_\Uparrow^R\Gamma_\Downarrow^L$.
\label{Fig1}}%
\end{figure}

Our measurements reveal a pronounced asymmetry in the step height of $G$ with respect to the polarity of $V$,
as recently predicted by Paaske {\em et al.}~\cite{Paaske}, in a model with a rather generic form of the SOI interaction.
The asymmetry is found to depend on the magnitude and direction of $B$.
Our results are consistent with an explanation based on the Luttinger Hamiltonian for the valence band of the semiconductor.
The spin selectivity of tunneling arises from an interplay of the complex structure of the valence band with the orbital effect of the magnetic field.
At $B=0$, the time-reversal symmetry ensures that the two states forming the Kramers doublet in the QD are indistinguishable
and the spin selectivity of tunneling vanishes.

\begin{figure}
\begin{center}
\begin{minipage}[t]{0.49\columnwidth}
\vspace{0pt}\includegraphics[width=1.0\textwidth]{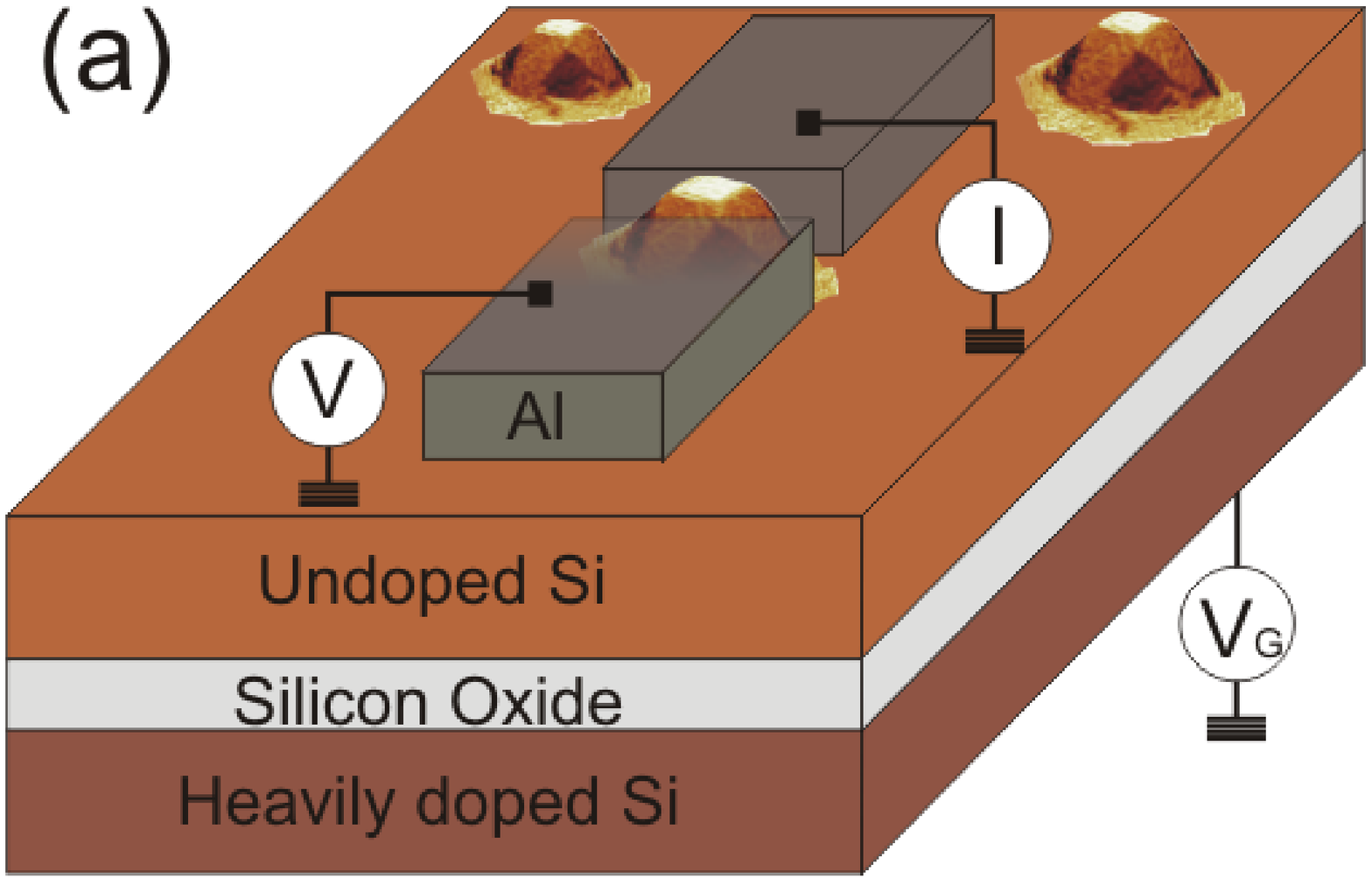}
\end{minipage}
\hfill
\begin{minipage}[t]{0.49\columnwidth}
\vspace{0pt}\includegraphics[width=1.0\textwidth]{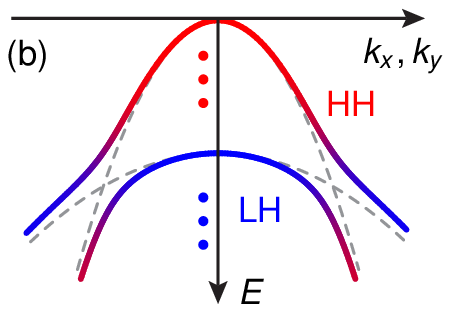}
\end{minipage}\par
\vspace{0pt}\includegraphics[width=\columnwidth]{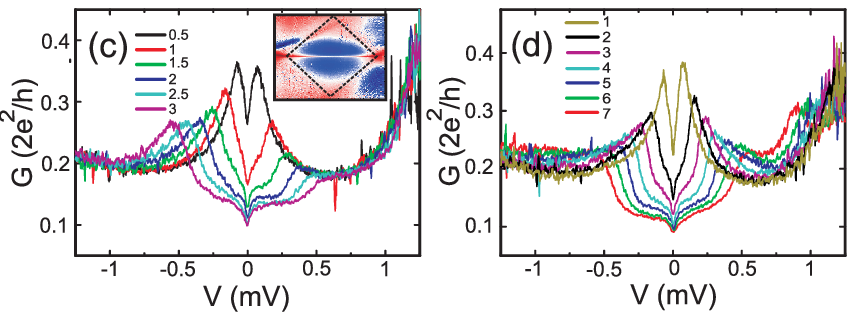}\par
\vspace{0pt}\includegraphics[width=\columnwidth]{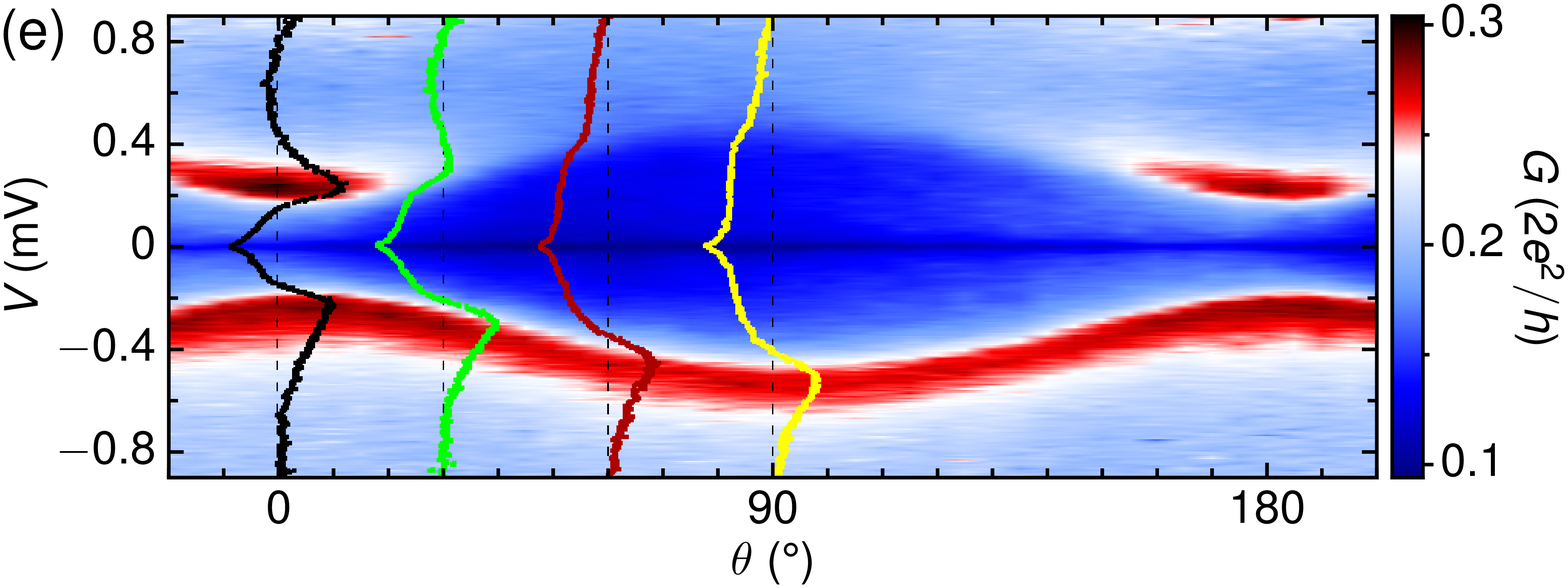}
\caption{
\label{Fig2}
(color online) (a) Schematic of a QD device fabricated from a SiGe self-assembled nanocrystal grown on a silicon-on-insulator substrate having a heavily doped
handle wafer which is used as a back gate~\cite{Katsaros}. (b)
Qualitative band diagram of a Ge-rich SiGe quantum well illustrating the effect of quantum confinement along the growth ($z$) direction:
HH and LH branches are split at $k_x = k_y = 0$ and anti-cross at finite $k_x$ or $k_y$.
The red dots indicate that many other HH subbands exist before the first LH subband is encountered.
(c) $G(V)$ for different perpendicular
$B$-fields from 0.5 to 3 T. Inset: $G(V_{G}, V)$ for a 75-mT perpendicular field
($V_{G}$ spans a range of $850\,{\rm mV}$ and $V$ ranges from -3.5 to 3.5 mV).
(d) $G(V)$ for different parallel fields from 1 to 7 T.
The Zeeman splitting of the Kondo peak is asymmetric in (c) and symmetric in (d).
(e) Angular dependence of the split Kondo peak for a fixed $V_{G}$ and $B=3\,{\rm T}$.
Superimposed $G(V)$ traces for $\theta = 0$, $30$, $60$, and $90$ degrees.}
\end{center}
\end{figure}

It is interesting to note that the transport characteristics of a QD with SOI coupled to normal leads
are similar to those of a QD without SOI coupled to FM leads.
We illustrate this similarity in Fig.~\ref{Fig1}, where we consider the simplest case, in which the Zeeman interaction
and the two spin-dependent tunnel contacts have collinear quantization directions.

We have studied the low-temperature magneto-transport properties of individual SiGe self-assembled QDs with
a base diameter $d \approx 80$ nm and a height $w \approx 20$ nm.
The hole motion is strongly quantized along the growth direction $[001]$.
A schematic of a typical QD contacted with Al electrodes is shown in Fig.~\ref{Fig2}(a).
For such QDs, the hole wave function is generally composed of both heavy holes (HHs) and light holes (LHs).
Due to the confinement and compressive strain, the degeneracy between the HH and LH branches, present in bulk at the $\Gamma$-point,
is lifted.
The HHs become energetically favorable.
In Fig.~\ref{Fig2}(b), we illustrate the interaction between a HH and a LH branch in the two-dimensional (2D) case.
The split-off band is far away in energy due to a large $\Delta_{\rm SO}$.
HHs and LHs are states of angular momentum $\frac{3}{2}$ with projections $\pm \frac{3}{2}$ and $\pm \frac{1}{2}$, respectively, see Appendix~\ref{secondtheory}.
We remark that, contrary to the HH states, the LH states cannot be factorized into a product of spin and orbital components.

The stability diagram, $G(V_G, V)$, of a QD device is shown in the inset of Fig.~\ref{Fig2}(c).
The diamond-shape region delimited by dashed lines highlights the CB regime for an odd number of confined holes.
While $G$ is generally suppressed within this CB diamond, a $G$ resonance can be identified at $V = 0$, providing a clear signature of a spin-$1/2$ Kondo effect~\cite{Goldhaber}.
At finite $B$, this resonance is split by the Zeeman effect as shown in Figs.~\ref{Fig2} (c) and (d) for perpendicular and parallel $B$, respectively
(all $G(V)$ traces were taken at the same $V_G$).

\begin{figure}
\includegraphics[width=8cm, keepaspectratio]{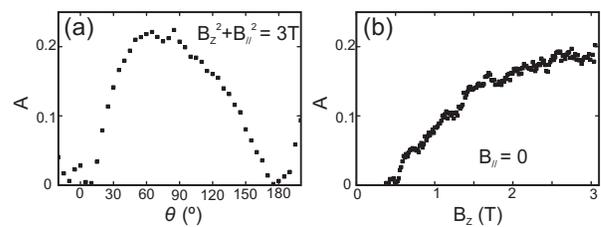}
\caption{\label{Fig3}
Asymmetry parameter $A$ as a function of (a) $\theta$ and (b) perpendicular $B$.
We note that, by subtracting the elastic CT contribution from $G(V)$, $A$ becomes larger than $0.5$.
}
\end{figure}

For perpendicular $B$ [Fig.~\ref{Fig2}(c)], the splitting of the Kondo peak is clearly asymmetric with respect to a sign change in $V$.
The asymmetry in $G$ arises at the onset of spin-flip inelastic CT (i.e. for $|eV| > E_Z$).
For parallel $B$, however, the asymmetry is practically absent [Fig.~\ref{Fig2}(d)].
To further investigate this anisotropy, a sequence of $G(V)$ traces was taken while rotating a $3\,{\rm T}$ field in a plane perpendicular to the substrate.
The resulting data, $G(\theta,V)$, are shown in Fig.~\ref{Fig2}(e), with $\theta$ being the angle between the field and the substrate plane.
Along with a variation in the Zeeman splitting of the Kondo peak, caused by the $\theta$-dependent hole g-factor~\cite{Katsaros},
the asymmetry becomes progressively more pronounced when going from $\theta = 0$ (or $180^\circ$) towards $\theta = 90^\circ$.

The asymmetry observed in $G(V)$ can be quantified by
$A = \left|\frac{G_{-}-G_{+}}{G_{-}+G_{+}}\right|$, where $G_{\pm} = G(\pm E_Z/e)$.
The detailed $A(\theta)$ dependence, extracted from Fig.~\ref{Fig2}(e), is shown in Fig.~\ref{Fig3}(a).
$A \approx 0$ for $\theta = 0$ (or $180^\circ$) and it increases monotonically up to $0.2$ for $\theta$ approaching $90^\circ$.
The same qualitative behavior was observed in a second device, which did not display Kondo effect, see Appendix~\ref{seconddevice}.
The asymmetry $A$ reaches $0.4$ at $3\,{\rm T}$ for that device.
We remark that,
although the first device shows larger conductance due to the Kondo effect,
the asymmetry $A$ in both devices is a consequence of spin-dependent tunnel rates.

In order to explain the microscopic origin of the measured effect,
we represent the Luttinger Hamiltonian~\cite{Luttinger}
as a block matrix in the basis of HH ($h$) and LH ($l$) states,
\begin{equation}
H=
\left(
\begin{array}{cc}
H_{hh} & H_{hl}\\
H_{lh} & H_{ll}
\end{array}
\right).
\label{eqHHhhHllHlhHhl}
\end{equation}
In the blocks $H_{hh}$ and $H_{ll}$,
we discard all terms that vanish in the 2D limit ($w/d\to 0$),
whereas in the blocks $H_{hl}$ and $H_{lh}$, we keep only the leading-order in $w/d$ terms.
A systematic expansion around the 2D limit is outlined in the Appendix~\ref{secondtheory}.
The blocks $H_{hh}$ and $H_{ll}$ assume a familiar form
\begin{eqnarray}
H_{hh/ll}&=&\frac{\gamma_1\pm\gamma_2}{2m}\left(k_x^2+k_y^2\right)+\frac{\gamma_1\mp 2\gamma_2}{2m}k_z^2
\nonumber\\
&&+\frac{1}{2}\mu_B\bm{\sigma}\cdot g_{h/l}\cdot\bm{B}+ U(x,y) + V_{h/l}(z),
\label{eqHhhoHll}
\end{eqnarray}
where the axes $x$, $y$, and $z$ point along
the main crystallographic directions,
with $z\equiv [001]$ being the direction of the strongest quantization.
After the expansion around the 2D limit,
the kinetic momentum operators $k_x$ and $k_y$
contain only the component $B_z$,
whereas $k_z$ is independent of $\bm{B}$.
In Eq.~(\ref{eqHhhoHll}) and below,
$\gamma_1$, $\gamma_2$, $\gamma_3$, $\kappa$, and $q$
denote the Luttinger parameters~\cite{Luttinger} and $m$ denotes the bare electron mass.
The Pauli matrices $\bm{\sigma}=(\sigma_x,\sigma_y,\sigma_z)$ represent the remaining
pseudo-spin degree of freedom in each block.
We choose the following pseudo-spin basis~\cite{note1}:
\begin{eqnarray}
\left|\uparrow\right\rangle_h&=&\left|3/2,-3/2\right\rangle, \quad
\left|\downarrow\right\rangle_h=\left|3/2,+3/2\right\rangle,\nonumber\\
\left|\uparrow\right\rangle_l&=&\left|3/2,+1/2\right\rangle, \quad
\left|\downarrow\right\rangle_l=\left|3/2,-1/2\right\rangle.
\label{basisps}
\end{eqnarray}
In the $(x,y,z)$-frame, the tensors of the g-factor are diagonal:
$g_h={\rm diag}\left(0,0,-6\kappa\right)$
and
$g_l={\rm diag}\left(4\kappa,4\kappa,2\kappa\right)$,
where we neglected, for simplicity, the terms proportional to the smallest Luttinger parameter $q$.
The minus sign in $(g_{h})_{zz}$ is due to our basis choice in Eq.~(\ref{basisps}).
In Eq.~(\ref{eqHhhoHll}), we included an in-plane confining potential $U(x,y)$.
The motion along $z$ is confined to an infinitely-deep square well,
with different offsets, $V_h$ and $V_l>V_h$, due to strain.

The blocks $H_{hl}$ and $H_{lh}$ are given by
\begin{equation}
H_{hl}= \left(H_{lh}\right)^\dagger=
i\frac{\sqrt{3}\gamma_3}{m}
\left( k_x\sigma_y + k_y\sigma_x\right)k_z.
\label{Hlh0order}
\end{equation}
These blocks intermix HHs and LHs, such that the wave
function of the hole in a given QD state assumes the general form $\Psi=\alpha\Psi_h+\beta\Psi_l$.
In terms of the true-spin states,
such a wave function consists of a superposition of the spin-up ($\uparrow$) and
spin-down ($\downarrow$) states entangled with the orbital degrees of freedom:
\begin{eqnarray}
\Psi_\Uparrow(\bm{r}) &=& \Phi_1(\bm{r})\uparrow + \chi_1(\bm{r})\downarrow,\nonumber\\
\Psi_\Downarrow(\bm{r}) &=& \chi_2(\bm{r})\uparrow + \Phi_2(\bm{r})\downarrow,
\label{eqPsiPsiKramers}
\end{eqnarray}
where $\Uparrow$ and $\Downarrow$ denote the components of the Kramers doublet in the QD.
Focusing on the first HH subband, we obtain by perturbation theory:
\begin{eqnarray}
\Phi_1(\bm{r})&=& \frac{\sqrt{2}\gamma_3}{m}\,{\cal U}_- k_-\psi_h(x,y)
\sum_n f^l_n(z)\frac{\left\langle f^l_n\right|k_z\left|f_1^h\right\rangle}{E_1^h-E_n^l},\nonumber\\
\chi_1(\bm{r})&=& {\cal U}_- \psi_h(x,y) f_1^h(z)
+\frac{2\gamma_3}{m} Z k_-\psi_h(x,y)\nonumber\\
&&\times
\sum_n f^l_n(z)\frac{\left\langle f^l_n\right|k_z\left|f_1^h\right\rangle}{E_1^h-E_n^l},
\label{eqPhichi1}
\end{eqnarray}
and similar expressions for $\Phi_2(\bm{r})$ and $\chi_2(\bm{r})$, obtained from
Eq.~(\ref{eqPhichi1}) by replacing ${\cal U}_-\to {\cal U}_+$, $k_-\to k_+$, and
$\psi_h(x,y)\to \psi_h^*(x,y)$.
In our notation,
${\cal U}_\pm =\mp\frac{1}{\sqrt{2}}\left(X\pm i Y\right)$ and
$k_\pm = \mp\frac{1}{\sqrt{2}}\left(k_x\pm i k_y\right)$.
The Bloch amplitudes $X$, $Y$, and $Z$ describe the valence band in the absence of SOI.
In blocks $H_{hh}$ and $H_{ll}$,
the motion along $z$ separates;
we denote the corresponding eigenenergies and eigenfunctions by
$E_n^{h/l}$ and $f^{h/l}_n(z)$, respectively.

The tunneling amplitudes $t^i_{\sigma s}$ are found as
\begin{eqnarray}
t^{i}_{\sigma s}=\sum_{u=X,Y,Z}T_u\left\langle u,\sigma|\Psi_s(\bm{r}_i)\right\rangle,
\label{tunamplTTT}
\end{eqnarray}
where $T_u$ is the coupling strength between Bloch amplitude $u$ and the lead,
and  $\left\langle u,\sigma|\Psi_s(\bm{r}_i)\right\rangle$
are the projections of the QD eigenstates $\Psi_s(\bm{r})$, see Eq.~(\ref{eqPsiPsiKramers}),
onto the product state of Bloch amplitude  $u$ and spinor $\left|\sigma\right\rangle$.
The tunneling amplitudes in Eq.~(\ref{tunamplTTT}) depend on the point of tunneling, $\bm{r}_i=\bm{r}_L,\bm{r}_R$,
the component of the true spin in the lead, $\sigma=\uparrow,\downarrow$,
and the component of the Kramers doublet on the dot, $s=\Uparrow,\Downarrow$.
We remark that $T_X$, $T_Y$, and $T_Z$ appear in Eq.~(\ref{tunamplTTT}) as phenomenological parameters.
They depend on the details of the metal-semiconductor interface and cannot be determined within the $\bm{k\cdot p}$-theory used here.
We find
\begin{equation}
t^i_{\sigma s}\propto\left(
\begin{array}{cc}
\bar{\Phi}_1(\bm{r}_i) & \bar{\chi}_2(\bm{r}_i) \\
\bar{\chi}_1(\bm{r}_i) & \bar{\Phi}_2(\bm{r}_i)
\end{array}
\right),
\label{eqtsgmsFchir}
\end{equation}
where
$\bar{\Phi}_i(\bm{r})$ and $\bar{\chi}_i(\bm{r})$ are obtained from Eq.~(\ref{eqPhichi1}) by replacing
$Z\to T_Z$ and ${\cal U}_\pm\to \mp\frac{1}{\sqrt{2}}\left(T_X\pm i T_Y\right)$.

\begin{figure}[t]
\begin{center}
\vspace{0pt}\includegraphics[width=0.32\columnwidth]{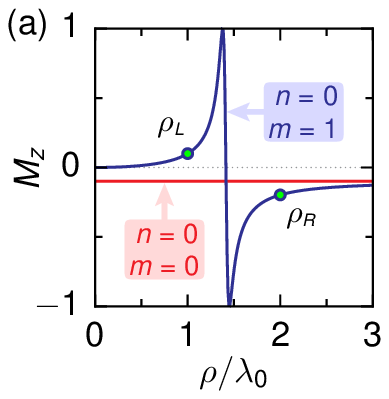}\hfill
\vspace{0pt}\includegraphics[width=0.32\columnwidth]{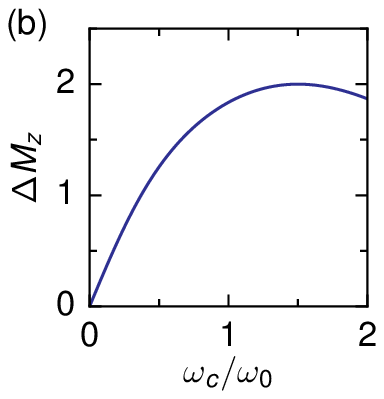}\hfill
\vspace{0pt}\includegraphics[width=0.34\columnwidth]{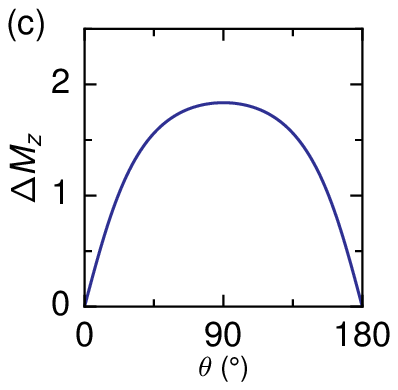}
\caption{
(color online) (a) Tunneling polarization $M_z$ as a function of the space coordinate $\rho$
for two Fock-Darwin states $(n, m)$ as indicated and for $\omega_c=0.1\omega_0$.
(b) $B$-field dependence of the difference $\Delta M_z=M_z(\rho_L) - M_z(\rho_R)$,
for the values of $\rho_L$ and $\rho_R$ indicated in (a).
(c) $\Delta M_z$ as a function of $\theta$ for a fixed value of $\left|\bm{B}\right|$.
The value of $\left|\bm{B}\right|$ corresponds to $\omega_c=\omega_0$ at $\theta = 90^\circ$.
\label{FigTh}
}
\end{center}
\end{figure}

The spin selectivity of the tunneling is best seen in
the matrix of the tunnel rates,
$\Gamma_{ss'}=\pi\sum_{\sigma}t^*_{\sigma s}\nu_{\sigma}t_{\sigma s'}$.
With $\nu_\uparrow=\nu_\downarrow$ (case of non-FM leads)	
we find, up to a common factor,
\begin{equation}
\left(
\begin{array}{cc}
\Gamma_{\Uparrow\Uparrow} &
\Gamma_{\Uparrow\Downarrow}\\
\Gamma_{\Downarrow\Uparrow}&
\Gamma_{\Downarrow\Downarrow}
\end{array}
\right)
\propto\left(
\begin{array}{cc}
\left|\bar{\Phi}_1\right|^2+\left|\bar{\chi}_1\right|^2 &
\bar{\Phi}_1^*\bar{\chi}_2+\bar{\chi}_1^*\bar{\Phi}_2\\
\bar{\Phi}_2^*\bar{\chi}_1+\bar{\chi}_2^*\bar{\Phi}_1
& \left|\bar{\Phi}_2\right|^2+\left|\bar{\chi}_2\right|^2
\end{array}
\right).
\label{eqttdaggerPhichi}
\end{equation}
At $B=0$, time-reversal symmetry requires that
\begin{equation}
\Phi_2(\bm{r})=\left[\Phi_1(\bm{r})\right]^*\quad\mbox{and}\quad\chi_2(\bm{r})=-\left[\chi_1(\bm{r})\right]^*,
\label{relations}
\end{equation}
leading to $\Gamma_{\Uparrow\Uparrow}=\Gamma_{\Downarrow\Downarrow}$ and $\Gamma_{\Uparrow\Downarrow}=\Gamma_{\Downarrow\Uparrow}=0$ in Eq.~(\ref{eqttdaggerPhichi}).

At $B\neq 0$, however, the orbital effect of the $B$-field modifies the functions $\Phi_i(\bm{r})$ and $\chi_i(\bm{r})$, such that the relations in Eq.~(\ref{relations}) are no longer satisfied.
In general, the matrix $\Gamma_{ss'}$ has nonzero off-diagonal elements.
Since it is a hermitian matrix, there exists a direction in space, $\bm{M}$, such that
an ${\rm SU}(2)$ rotation of the Kramers doublet by an angle $\angle zM$ makes the rate matrix diagonal,
$\Gamma = {\rm diag}(\Gamma_{\Uparrow},\Gamma_{\Downarrow})$, with $\Gamma_{\Uparrow}\geq \Gamma_{\Downarrow}$.
To quantify the spin selectivity of the tunneling, we define
\begin{equation}
\left|\bm{M}\right|=\frac{\Gamma_{\Uparrow} - \Gamma_{\Downarrow}}{\Gamma_{\Uparrow} + \Gamma_{\Downarrow}}.
\label{eqdefM}
\end{equation}
In respect to transport, $\bm{M}$ is analogous to the polarization vector of the FM.
Indeed, the maximum of spin selectivity in tunneling from a FM is achieved when the FM
is a half-metal, {\em e.g.}, $\nu_\uparrow\neq 0$ and $\nu_\downarrow = 0$.
This extreme case corresponds to $M=1$ and can be approached in our case by increasing $B_z$.

In order to illustrate the origin of the spin selectivity, we focus on the special case:
$T_X=T_Y=0$ and $T_Z\neq 0$ and refer to this tunneling model as the $Z$-model.
In the $Z$-model, vector $\bm{M}$ is parallel to the $z$-axis.
Tunneling to the hole states is possible only due to the admixture of the LH subbands.
Furthermore, in this model, the spin selectivity is determined by the fact that
$\bar{\chi}_1(\bm{r})\propto k_-\psi_h(x,y)$ and $\bar{\chi}_2(\bm{r})\propto k_+\psi_h^*(x,y)$, whereas $\bar{\Phi}_i(\bm{r})\equiv 0$.
Using this information in Eqs.~(\ref{eqttdaggerPhichi}) and~(\ref{eqdefM}), we specify $\psi_h(x,y)$ to the Fock-Darwin states~\cite{FockDarwin}.
Therefore, we assume that $U(x,y)$ in Eq.~(\ref{eqHhhoHll}) is given by $U(\rho)=m^*\omega_0^2\rho^2/2$,
where $m^*$ is the effective mass for in-plane motion, $\omega_0$ is the oscillator frequency of the harmonic potential,
and $\rho^2=x^2+y^2$.
For the first two states ($n=0$ and $m=0,-1$), we obtain
\begin{eqnarray}
M_z=-\frac{\omega\omega_c}{\omega^2+\omega_c^2/4},
\end{eqnarray}
where $\omega=\sqrt{\omega_0^2+\omega_c^2/4}$, and $\omega_c=eB_z/m^*c$.
For these states, $M_z$ depends on $B_z$ but not on $\rho$, see Fig.~\ref{FigTh}(a).
For $B_z \neq 0$, the contacts will exhibit spin-dependent tunnel rates with the same polarization value $M_z$ regardless of the point-tunneling position.
In such a case, no asymmetry in the inelastic CT is expected.

The situation changes starting from $n=0$ and $m=1$, where
\begin{eqnarray}
M_z=\left[
\frac{\omega}{\omega_c}f(\rho)+\frac{\omega_c}{4\omega}\frac{1}{f(\rho)}\right]^{-1},
\label{eqMz}
\end{eqnarray}
with $f(\rho)=2\hbar/(m^*\omega\rho^2) -1$.
Now $M_z$ depends both on $B_z$ and on $\rho$, see Fig.~\ref{FigTh}(a).
The spin polarization of two contacts positioned arbitrarily on a QD may differ significantly from each other,
see, {\em e.g.}, points $\rho_L$ and $\rho_R$ in Fig.~\ref{FigTh}(a).
The asymmetry in the inelastic CT is related to $\Delta M_z=M_z(\rho_L) - M_z(\rho_R)\neq 0$.
$\Delta M_z$ increases with $B_z$ [Fig.~\ref{FigTh} (b)], displaying at the same time strong dependence on the $B$-field direction [Fig.~\ref{FigTh} (c)],
in good qualitative agreement with the results in Fig.~\ref{Fig3}.

The described joint effect of SOI and Zeeman splitting explains our experimental findings.
In addition, it opens the door to an original scheme for measuring Rabi spin oscillations in hole confinement QDs.
Let us consider a spin-1/2 QD in the CB regime under a perpendicular $B$ of the order of a few T.
In such a case, a transport characteristic of the type shown in Fig. 2(c) is to be expected.
For $V=0$, no current flows through the QD. Yet we suggest that a finite current could be generated by a resonant rf field (at frequency $f=E_Z/h$)
capable of inducing coherent oscillations between the Zeeman-split states of the QD.
In fact, as the excited $\Uparrow$ state becomes populated it can decay to the ground $\Downarrow$ state by an inelastic CT
process---a hole tunnels out of the QD from the $\Uparrow$ state being replaced by another hole tunneling into the $\Downarrow$ state.
Because $\Uparrow$ and $\Downarrow$ states have tunnel couplings with opposite asymmetries, in a configuration such as the one depicted in Fig. 1(b) the most favorable CT relaxation process would involve the transfer of a hole from the right to the left contact.  Hence a net dc current could be driven by a continuous resonant irradiation.
In addition, combining rf bursts with syncronized $V_G$ pulses may enable the coherent control of the QD pseudo-spin states. In this scheme, well-defined pseudo-spin rotations would be performed in the deep CB regime (i.e. during a negative $V_G$ pulse), whereas pseudo-spin read-out would take place in the CT regime.

We acknowledge J.~Paaske for helpful discussions and
A.~Rastelli and H.~von~Kaenel for providing the STM image used in Fig.~\ref{Fig1}(a).
The work was supported by the Agence Nationale de la Recherche (through the ACCESS and COHESION projects),
US DOE Contract No. DE-FG02-08ER46482 (Yale),
and the Nanosciences Foundation at Grenoble, France.
G.K. acknowledges support from the Deutsche Forschungsgemeinschaft.

\appendix

\begin{figure}
\vspace{0pt}\includegraphics[width=8.6cm, keepaspectratio]{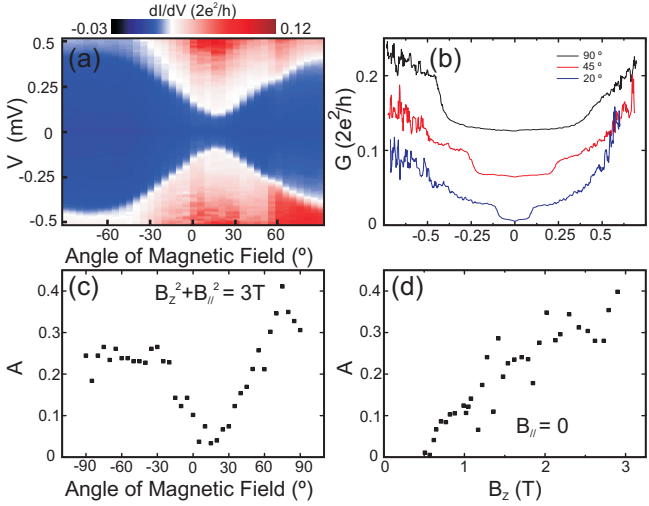}
\caption{\label{sup}
(a) Evolution of the differential conductance  $dI/dV$ vs the angle of magnetic field and $V$. The amplitude of the magnetic field is fixed to 3T. (b) Characteristic traces of $dI/dV$ vs
$V$ for $20$ (blue), $45$ (red)and $90$ (black) degrees,
respectively. The traces have been shifted by $0.06 x 2e^2/h$ for clarity. (c) Plot showing the evolution of the asymmetry $A$ vs the angle of the magnetic field. (d) Plot showing the evolution of $A$ vs the value of the perpendicular field.}
\end{figure}

\section{Second device}\label{seconddevice}

Similar asymmetries as the ones described in the main text were observed also for a second device, less strongly coupled to the metallic leads. Figure~\ref{sup} (a) is a plot of the differential conductance $G$ versus the angle of magnetic field and the bias voltage $V$ for a fixed value of 3 T. Differently to the first device the minimum g-factor is not observed for a parallel magnetic field but there is a shift by about 15-20 degrees. Some characteristic traces of G vs $V$ taken at 20 (blue), 45 (red) and 90 (black) degrees
are shown in Fig.~\ref{sup} (b).
Interestingly, the position of the minimum g-factor coincides with the case for which no asymmetry is appearing (15-20 degrees).
We believe that both observations are because the 2D plane of the wave function is not parallel to the substrate.

The asymmetry $A$ follows the same trends as were observed in the first device. It is almost zero for 15-20 degrees (the position of the minimum $g_\parallel$) and it obtains values of about 0.35-0.4 at large out of plane angles of the magnetic field [Fig.~\ref{sup} (c)]. The difference in the value of A for positive and negative magnetic fields is attributed again to the different angle the 2D hole wavefunction plane forms with the magnetic field. Finally, Fig.~\ref{sup} (d), verifies that the asymmetry increases with $B_z$.

We remark that in the absence of misalignment
the asymmetry $A$ obeys the relation
\begin{equation}
A(B_z)=A(-B_z),
\label{eqAAbzmBz}
\end{equation}
which holds within the experimental accuracy for the device described in the main text.
Equation~(\ref{eqAAbzmBz}) can be understood as follows.
On the one hand, the spin-selective part of $\Gamma_{ss'}$ is proportional to the orbital $B_z$
and therefore it changes sign when flipping the direction of the magnetic field.
On the other hand, the Zeeman energy is also changing sign when flipping the direction of the magnetic field,
swapping thus the roles of the ground and the excited state.
As a result, the measured cotunneling asymmetry $A$ does not change sign when changing $\bm{B}\to-\bm{B}$.

\section{Expansion around the 2D limit}\label{secondtheory}
Our starting point is the Luttinger Hamiltonian~\cite{Luttinger1},
\begin{eqnarray}
&&H = \frac{1}{2m}\left(\gamma_1 +\frac{5}{2}\gamma_2\right)k^2
-\frac{\gamma_2}{m}\left(k_x^2J_x^2+k_y^2J_y^2+k_z^2J_z^2\right)\nonumber\\
&&-\frac{2\gamma_3}{m}\left(\{k_xk_y\}\{J_xJ_y\}+\{k_yk_z\}\{J_yJ_z\}+\{k_zk_x\}\{J_zJ_x\}\right)\nonumber\\
&&+\frac{e\hbar}{mc}\kappa\,\bm{J\cdot B}+\frac{e\hbar}{mc}q\left(J_x^3B_x+J_y^3B_y+J_z^3B_z\right),
\label{HamLutt}
\end{eqnarray}
where $m$ is the mass of the electron in vacuum,
$\gamma_1$, $\gamma_2$, $\gamma_3$, $\kappa$, and $q$ are the Luttinger parameters,
$\bm{k}=(k_x,k_y,k_z)$ is the momentum of the band electron,
\begin{equation}
\bm{k}=-i\hbar\frac{\partial}{\partial \bm{r}} +\frac{e}{c}\bm{A}(\bm{r}),
\end{equation}
with $\bm{A}(\bm{r})$ being the vector potential due to the magnetic field,
$e$ the elementary charge ($e>0$), and $c$ the speed of light.
Further, $\{\dots\}$ denotes the symmetrized product, e.g.\ %
\begin{equation}
\{k_xk_y\} =\frac{1}{2}\left(k_xk_y+k_yk_x\right),
\end{equation}
and $\bm{J}=(J_x,J_y,J_z)$ are $4\times 4$ matrices
representing the spin $J=3/2$ in a basis of choice.
We choose the basis~\cite{Abakumov}
\begin{equation}
\left\{
\left|\frac{3}{2},+\frac{3}{2}\right\rangle,\quad
\left|\frac{3}{2},+\frac{1}{2}\right\rangle,\quad
\left|\frac{3}{2},-\frac{1}{2}\right\rangle,\quad
\left|\frac{3}{2},-\frac{3}{2}\right\rangle
\right\}
\label{eqbasis32}
\end{equation}
where
\begin{eqnarray}
\left|\frac{3}{2},+\frac{3}{2}\right\rangle&=&-\frac{1}{\sqrt{2}}\left(X+iY\right)\uparrow,\nonumber\\
\left|\frac{3}{2},-\frac{3}{2}\right\rangle&=&\frac{1}{\sqrt{2}}\left(X-iY\right)\downarrow,\nonumber\\
\left|\frac{3}{2},+\frac{1}{2}\right\rangle&=&\frac{1}{\sqrt{6}}\left[-\left(X+iY\right)\downarrow+2Z\uparrow\right],\nonumber\\
\left|\frac{3}{2},-\frac{1}{2}\right\rangle&=&\frac{1}{\sqrt{6}}\left[\left(X-iY\right)\uparrow+2Z\downarrow\right].
\label{hhllbasis0}
\end{eqnarray}
Here, the Bloch amplitudes $X$, $Y$, and $Z$ are chosen to be real.
They belong to the representation $\Gamma_{25'}$ of the group $O_h$
and transform under the group operations as
$X\equiv yz$, $Y\equiv xz$, and $Z\equiv xy$~\cite{DKK,CardonaPollak,YuCardona}.
The basis functions $X$, $Y$, and $Z$ form a subspace that
is equivalent to the space of the angular momentum $I=1$~\cite{Luttinger1}.
The states in Eq.~(\ref{hhllbasis0}) originate from the addition~\cite{LK}
$\bm{J}=\bm{I}+\bm{S}$, where $\bm{S}$ is the electron spin ($S=1/2$)
in the usual basis $\{\uparrow,\downarrow\}$.
The subspace of  $J=3/2$ is shown in Eq.~(\ref{hhllbasis0}),
whereas the subspace of $J=1/2$ is neglected, because it corresponds to the
split-off band, i.e.\ the band that is shifted, due to the spin-orbit interaction,
by an amount $\Delta_{\rm SO}$ below the top of the valence band.
The Luttinger Hamiltonian describes the very top of the valance band, at energies
$E\ll\Delta_{\rm SO}$.
In the basis given by Eqs.~(\ref{eqbasis32}) and~(\ref{hhllbasis0}),
the matrices of $J=3/2$ read~\cite{Abakumov}
\begin{equation}
J_x=\left(
\begin{array}{cccc}
0&\frac{\sqrt{3}}{2}&0&0\\
\frac{\sqrt{3}}{2}&0&1&0\\
0&1&0&\frac{\sqrt{3}}{2}\\
0&0&\frac{\sqrt{3}}{2}&0\\
\end{array}
\right),
\end{equation}
\begin{equation}
J_y=\left(
\begin{array}{cccc}
0&-i\frac{\sqrt{3}}{2}&0&0\\
i\frac{\sqrt{3}}{2}&0&-i&0\\
0&i&0&-i\frac{\sqrt{3}}{2}\\
0&0&i\frac{\sqrt{3}}{2}&0\\
\end{array}
\right),
\end{equation}
\begin{equation}
J_z=\left(
\begin{array}{cccc}
\frac{3}{2}&0&0&0\\
0&\frac{1}{2}&0&0\\
0&0&-\frac{1}{2}&0\\
0&0&0&-\frac{3}{2}\\
\end{array}
\right).
\end{equation}
In Eq.~(\ref{HamLutt}), the axes $x$, $y$, and $z$ are fixed along the main crystallographic directions of the cubic crystal.
We choose the axis $z$ to point along the growth direction of the nanocrystal, making it the axis of the strongest size quantization.
The idea of expanding the Luttinger Hamiltonian around the 2D limit consists in
regarding quantities like $z$ and $\partial/\partial z$ as proportional to $w$ and $1/w$, respectively.
Here, $w$ is the width of the 2D layer (i.e.\ height of nanocrystal),
which is considered to be much smaller than the nanocrystal diameter $d$.
On the other hand, quantities like $x$ and $\partial/\partial x$ are regarded as proportional to
$d$ and $1/d$, respectively~\cite{note1}.
Before expanding in powers of $w/d\ll 1$,
it is convenient to represent the Luttinger Hamiltonian in a block form.

We use two projection operators, $p_h$ and $p_l$, which project on the subspaces of the heavy ($h$) and light ($l$) holes.
In terms of $J_z$, they are written as
\begin{eqnarray}
p_h&=&\frac{1}{2}\left(J_z^2-\frac{1}{4}\right),\nonumber\\
p_l&=&\frac{1}{2}\left(\frac{9}{4}-J_z^2\right).
\end{eqnarray}
$p_h$ and $p_l$ resolve the unity, $p_h+p_l=1$,
and have the usual properties of projection operators: $p_h^2=p_h$, $p_l^2=p_l$, and $p_hp_l=p_lp_h=0$.
The Luttinger Hamiltonian in Eq.~(\ref{HamLutt}) can then be written as follows
\begin{eqnarray}
H&=& (p_h+p_l)H(p_h+p_l)=
H_{hh}\left|h\right\rangle\left\langle h\right|+ H_{hl}\left|h\right\rangle\left\langle l\right| \nonumber\\
&+& H_{lh}\left|l\right\rangle\left\langle h\right|+H_{ll}\left|l\right\rangle\left\langle l\right|,
\end{eqnarray}
which makes up a matrix in the $(h, l)$-space,
\begin{equation}
H=
\left(
\begin{array}{cc}
H_{hh} & H_{hl}\\
H_{lh} & H_{ll}
\end{array}
\right).
\end{equation}
Each element $H_{ij}$ can be represented as
a $2\times 2$ matrix in the space of the pseudo-spin.
The blocks on the diagonal read
\begin{eqnarray}
H_{hh}&=&\frac{\gamma_1+\gamma_2}{2m}\left(k_x^2+k_y^2\right)+\frac{\gamma_1-2\gamma_2}{2m}k_z^2
\nonumber\\
&&+\frac{1}{2}\mu_B\bm{\sigma}\cdot g_h\cdot\bm{B},\nonumber\\
H_{ll}&=&\frac{\gamma_1-\gamma_2}{2m}\left(k_x^2+k_y^2\right)+\frac{\gamma_1+2\gamma_2}{2m}k_z^2
\nonumber\\
&&+\frac{1}{2}\mu_B\bm{\sigma}\cdot g_l\cdot\bm{B},
\end{eqnarray}
where the g-factors $g_h$ and $g_l$ are tensors.
In the frame $(x,y,z)$, they are diagonal:
\begin{equation}
g_h=
\left(
\begin{array}{ccc}
3q & 0 & 0 \\
0 & 3q &0 \\
0 & 0 & -6\kappa -\frac{27}{2}q
\end{array}
\right),
\end{equation}
and
\begin{equation}
g_l=
\left(
\begin{array}{ccc}
4\kappa+10q & 0 & 0 \\
0 & 4\kappa+10q &0 \\
0 & 0 & 2\kappa +\frac{1}{2}q
\end{array}
\right).
\end{equation}
The off-blocks are related to each other by hermiticity,
\begin{equation}
H_{hl}=\left(H_{lh}\right)^\dagger.
\end{equation}
For $H_{lh}$, we have
\begin{eqnarray}
H_{lh}&=&
-i\frac{\sqrt{3}\gamma_3}{m}\left(\left\{k_xk_z\right\}\sigma_y+\left\{k_yk_z\right\}\sigma_x\right)\nonumber\\
&&
-\frac{\sqrt{3}\gamma_2}{2m}\left(k_x^2-k_y^2\right)+
i\frac{\sqrt{3}\gamma_3}{m}\left\{k_xk_y\right\}\sigma_z\nonumber\\
&&+\sqrt{3}\mu_B\left(\kappa+\frac{7}{4}q\right)\left(\sigma_xB_x-\sigma_yB_y\right).
\label{LuttHlhexct}
\end{eqnarray}
All we have done so far was to rewrite the Luttinger Hamiltonian in a block form.
Next we proceed with the expansion in powers of $w/d\ll 1$ as explained above.

We allow for gauges of the form
\begin{eqnarray}
A_x&=&zB_y -\tau y B_z,\nonumber\\
A_y&=&-zB_x+ (1-\tau)xB_z,\nonumber\\
A_z&=&0,
\end{eqnarray}
where $\tau$ is a real number expressing the remaining gauge freedom in two dimensions.
After taking the 2D limit, we will be able to use a reduced (2D) vector potential,
$\bm{a}=(a_x,a_y)$,
which is given by
\begin{eqnarray}
a_x&=& -\tau y B_z,\nonumber\\
a_y&=& (1-\tau)xB_z.
\label{eqaxaytau}
\end{eqnarray}
Having in mind such a transition, we pull out the $z$-dependence from $k_x$ and $k_y$,
\begin{eqnarray}
k_x&\to& k_x+ \frac{eB_y}{c}z,\nonumber\\
k_y&\to& k_y- \frac{eB_x}{c}z.
\label{to2D}
\end{eqnarray}
Here, on the right-hand side, $k_x$ and $k_y$ do not depend on $z$ anymore,
because they are given in terms of the 2D vector potential $\bm{a}(x,y)$ as
\begin{eqnarray}
k_x=-i\hbar\frac{\partial}{\partial x}+\frac{e}{c}a_x,\nonumber\\
k_x=-i\hbar\frac{\partial}{\partial y}+\frac{e}{c}a_y.
\end{eqnarray}
The next step is to substitute Eq.~(\ref{to2D}) into the Luttinger Hamiltonian
and to group the terms according to their order of $w/d$.

The substitution of Eq.~(\ref{to2D}) in the blocks $H_{hh}$ and $H_{ll}$
produces linear in $k_x$ and $k_y$ terms which are not multiplied by any Pauli matrix.
Such terms can be gauged away after integration over $z$, since they correspond to a constant shift in
$a_x$ and $a_y$.
They also admix higher heavy-hole subbands and slightly renormalize the inplane mass,
but this admixture, as well as the mass renormalization, vanishes in the limit $w\to 0$,
because the corresponding perturbation is proportional to $z$.
Therefore, we dispense with the new terms generated in $H_{hh}$ and $H_{ll}$.

We note that, for the blocks $H_{hh}$ and $H_{ll}$, the transition to 2D
is identical to what is usually done for electrons in the conduction band.

For the blocks $H_{lh}$ and $H_{hl}$, we make the substitution in Eq.~(\ref{to2D}) and obtain lots of terms.
The origin of each term can be traced back
through the following intermediate step:
\begin{eqnarray}
\left\{k_x k_z\right\} &\to & k_x k_z +\frac{eB_y}{c}\left\{z k_z\right\},\nonumber\\
\left\{k_y k_z\right\} &\to & k_y k_z -\frac{eB_x}{c}\left\{z k_z\right\},\nonumber\\
k_x^2-k_y^2 &\to & k_x^2-k_y^2 +\frac{2e}{c}\left(k_xB_y+k_yB_x\right)z \nonumber\\
&&+\frac{e^2}{c^2}\left(B_y^2-B_x^2\right)z^2,\nonumber\\
\left\{k_x k_y\right\} &\to & \left\{k_x k_y\right\} +\frac{e}{c}\left(k_yB_y-k_xB_x\right)z\nonumber\\
&&-\frac{e^2}{c^2}B_xB_yz^2.
\label{eqrepintstep}
\end{eqnarray}
Then, we classify all terms according to their order of $w/d\ll 1$.

The leading order is that of $(w/d)^{-1}\gg1$ and the off-block acquires the following main term
\begin{equation}
H_{lh}^{(0)} =
-i\frac{\sqrt{3}\gamma_3}{m}
\left( k_x\sigma_y + k_y\sigma_x\right)k_z.
\label{Hlh0order}
\end{equation}
It is important to remark that $k_x$ and $k_y$ contain only the $z$-component of the magnetic field.
The transverse components $B_x$ and $B_y$ do not appear in $H_{lh}$
at this order of $w/d$.
On this reason, the component $B_z$ has a larger effect on breaking the time-reversal symmetry than the other two components.

The next order is that of $(w/d)^{0}\sim 1$ and the off-block acquires the following correction
\begin{eqnarray}
H_{lh}^{(1)}&=&
-\frac{\sqrt{3}\gamma_2}{2m}\left(k_x^2-k_y^2\right)+
i\frac{\sqrt{3}\gamma_3}{m}\left\{k_xk_y\right\}\sigma_z\nonumber\\
&&+\sqrt{3}\mu_B\left(\kappa+\frac{7}{4}q+\frac{2i\gamma_3}{\hbar}\left\{zk_z\right\}\right)\left(\sigma_xB_x-\sigma_yB_y\right).\nonumber\\
\label{LuttHlh1}
\end{eqnarray}

Further, there are two more orders: $(w/d)^{1}\ll 1$ and $(w/d)^{2}\ll 1$, originating from terms containing $z$ and $z^2$, respectively.
If they are required,
one can find them by substituting Eq.~(\ref{eqrepintstep}) into Eq.~(\ref{LuttHlhexct}).


\begin{thebibliography}{99}
\bibitem{note1}
The advantage of this choice is that the pseudo-spin transforms under the time reversal
as a spin $1/2$, i.e. $\left|\uparrow\right\rangle\to \left|\downarrow\right\rangle$ and $\left|\downarrow\right\rangle\to -\left|\uparrow\right\rangle$.
\bibitem{note1}
The same premise is made for $y$ and $\partial/\partial y$.
\end{thebibliography}
\end{document}